\documentclass[
    aps,
    pra,
    longbibliography,
    superscriptaddress,amsmath,amssymb,amsfonts, 
    tightenlines,
    twocolumn,
    ]{revtex4-1}
    
\usepackage[utf8]{inputenc}  
\usepackage[T1]{fontenc}     
\usepackage[english]{babel}  
\usepackage{hyperref}  
\usepackage{graphicx} 
\usepackage[babel]{microtype}  
\usepackage{amsmath,amssymb,amsthm,bm,amsfonts,mathrsfs,bbm} 
\usepackage{setspace}
\usepackage{times}
\usepackage{braket}
\usepackage[bottom]{footmisc}
\usepackage{blochsphere}
\usepackage{pgfplots}
\usepackage{ulem}
\usepackage{tcolorbox}
 
\pgfplotsset{compat=1.17}
\usetikzlibrary{calc,3d,shapes, pgfplots.external, intersections}
\usepackage{algorithm}
\usepackage{listings}
\usepackage{physics}
\usepackage[noend]{algorithmic}
\usepackage{eqparbox}

\usepackage{physics}
\usepackage{tikz}
\usepackage{mathtools}
\usetikzlibrary{arrows.meta}
\usepackage{csquotes}

\usepackage{xspace}  
\usepackage{pgfplots}
\usepackage{verbatim}
\usepackage{amsmath}

\newcommand\id{\leavevmode\hbox{\small1\kern-3.3pt\normalsize1}}

\usepackage{hyperref}

\usepackage{xcolor}
\usepackage{pdfpages}
\usepackage{pgffor}
\makeatletter
\AtBeginDocument{\let\LS@rot\@undefined}
\makeatother

\definecolor{backcolour}{rgb}{0.95,0.95,0.92}

\lstdefinestyle{mystyle}{
    backgroundcolor=\color{backcolour},   
    basicstyle=\ttfamily\footnotesize,
    breakatwhitespace=false,         
    breaklines=true,                 
    captionpos=b,                    
    keepspaces=true,                 
    numbers=none,                    
    numbersep=5pt,                  
    showspaces=false,                
    showstringspaces=false,
    showtabs=false,                  
    tabsize=2
}

\begin{document}

\title{GroverGPT: A Large Language Model with 8 Billion Parameters for Quantum Searching}

\author{Haoran Wang{$^*$}}
\affiliation{Department of Computer Science, The University of North Carolina at Chapel Hill, Chapel Hill, NC 27599, USA}

\author{Pingzhi Li{$^*$}}
\affiliation{Department of Computer Science, The University of North Carolina at Chapel Hill, Chapel Hill, NC 27599, USA}

\author{Min Chen}
\affiliation{Department of Computer Science, The University of Pittsburgh, Pittsburgh, PA 15260, USA}

\author{Jinglei Cheng}
\affiliation{Department of Computer Science, The University of Pittsburgh, Pittsburgh, PA 15260, USA}

\author{Junyu Liu{$^\ddagger$}}
\affiliation{Department of Computer Science, The University of Pittsburgh, Pittsburgh, PA 15260, USA}

\author{Tianlong Chen{$^\ddagger$}}
\affiliation{Department of Computer Science, The University of North Carolina at Chapel Hill, Chapel Hill, NC 27599, USA}

\maketitle

\noindent {$^{*}$These authors contributed equally to this work.}\\
\noindent {$^{\ddagger}$Co-corresponding authors.}\\
\href{mailto:tianlong@cs.unc.edu}{tianlong@cs.unc.edu},  \href{mailto:junyuliu@pitt.edu}{junyuliu@pitt.edu}
\\
\\
\noindent{\bf Quantum computing is an exciting non-Von Neumann paradigm, offering provable speedups over classical computing for specific problems. However, the practical limits of classical simulatability for quantum circuits remain unclear, especially with current noisy quantum devices. In this work, we explore the potential of leveraging Large Language Models (LLMs) to simulate the output of a quantum Turing machine using Grover's quantum circuits, known to provide quadratic speedups over classical counterparts. To this end, we developed GroverGPT, a specialized model based on LLaMA's 8-billion-parameter architecture, trained on over 15 trillion tokens. Unlike brute-force state-vector simulations, which demand substantial computational resources, GroverGPT employs pattern recognition to approximate quantum search algorithms without explicitly representing quantum states. Analyzing 97K quantum search instances, GroverGPT consistently outperformed OpenAI's GPT-4o (45\% accuracy), achieving nearly 100\% accuracy on 6- and 10-qubit datasets when trained on 4-qubit or larger datasets. It also demonstrated strong generalization, surpassing 95\% accuracy for systems with over 20 qubits when trained on 3- to 6-qubit data. Analysis indicates GroverGPT captures quantum features of Grover's search rather than classical patterns, supported by novel prompting strategies to enhance performance. Although accuracy declines with increasing system size, these findings offer insights into the practical boundaries of classical simulatability. This work suggests task-specific LLMs can surpass general-purpose models like GPT-4o in quantum algorithm learning and serve as powerful tools for advancing quantum research.}

\section{Introduction}\label{sec:intro}

Quantum computing harnesses fundamental quantum mechanical phenomena, such as superposition and entanglement, to solve certain computational problems exponentially faster than classical computers~\cite{shor1994algorithms,grover1996fast}. For instance, Grover's algorithm~\cite{grover1996fast}, a quantum algorithm for database searching, is proven to achieve a quadratic speedup over all known classical counterparts. However, the boundaries of \textit{quantum advantage}—where quantum algorithms outperform all classical counterparts for a specific task—remain unclear. In practice, a closely related question concerns classical simulability: if a quantum circuit can be efficiently simulated on classical computers, it is unlikely to demonstrate a significant advantage~\cite{nielsen2010quantum}.

Based on the construction of a quantum Turing machine, brute-force classical simulation—commonly referred to as state-vector simulation—incurs exponentially high memory costs for general quantum computing tasks. Smarter, approximate approaches, such as tensor networks, may perform better in practice~\cite{vidal2003efficient}, although they can still face exponentially high costs or exponentially large errors as the number of qubits scales for general tasks. The situation becomes even more complex with NISQ (Noisy Intermediate-Scale Quantum) devices~\cite{preskill2018quantum}; it remains unclear whether existing noisy, near-term commercial quantum computers can be classically simulated for commercially valuable problems~\cite{FASQ}.

Thus, the practical frontier of classical simulability can only be approached with \textit{large-scale high-performance computing}. For example, several claims of demonstrating quantum advantages in sampling algorithms~\cite{arute2019quantum} have been challenged by tensor network methods implemented on classical devices~\cite{oh2024classical,pan2022solving}. On the other hand, novel approaches utilizing Large Language Models (LLM) \cite{touvron2023llama,dubey2024llama,vavekanand2024llama}, one of the most powerful large-scale AI tools, present new possibilities for simulating quantum circuits \cite{liang2023unleashing,yang2024qcircuitnet,yao2024shadowgpt,fitzek2024rydberggpt}. 
Although notable progress has been achieved, these early efforts have primarily concentrated on prompt engineering using commercial models from OpenAI and similar providers, building open-source datasets, or developing toy models with only a few thousand parameters—significantly smaller than modern industrial-level models, which typically have at least a few billion parameters.

\begin{figure}[t]
    \centering
    \includegraphics[width=0.9\linewidth]{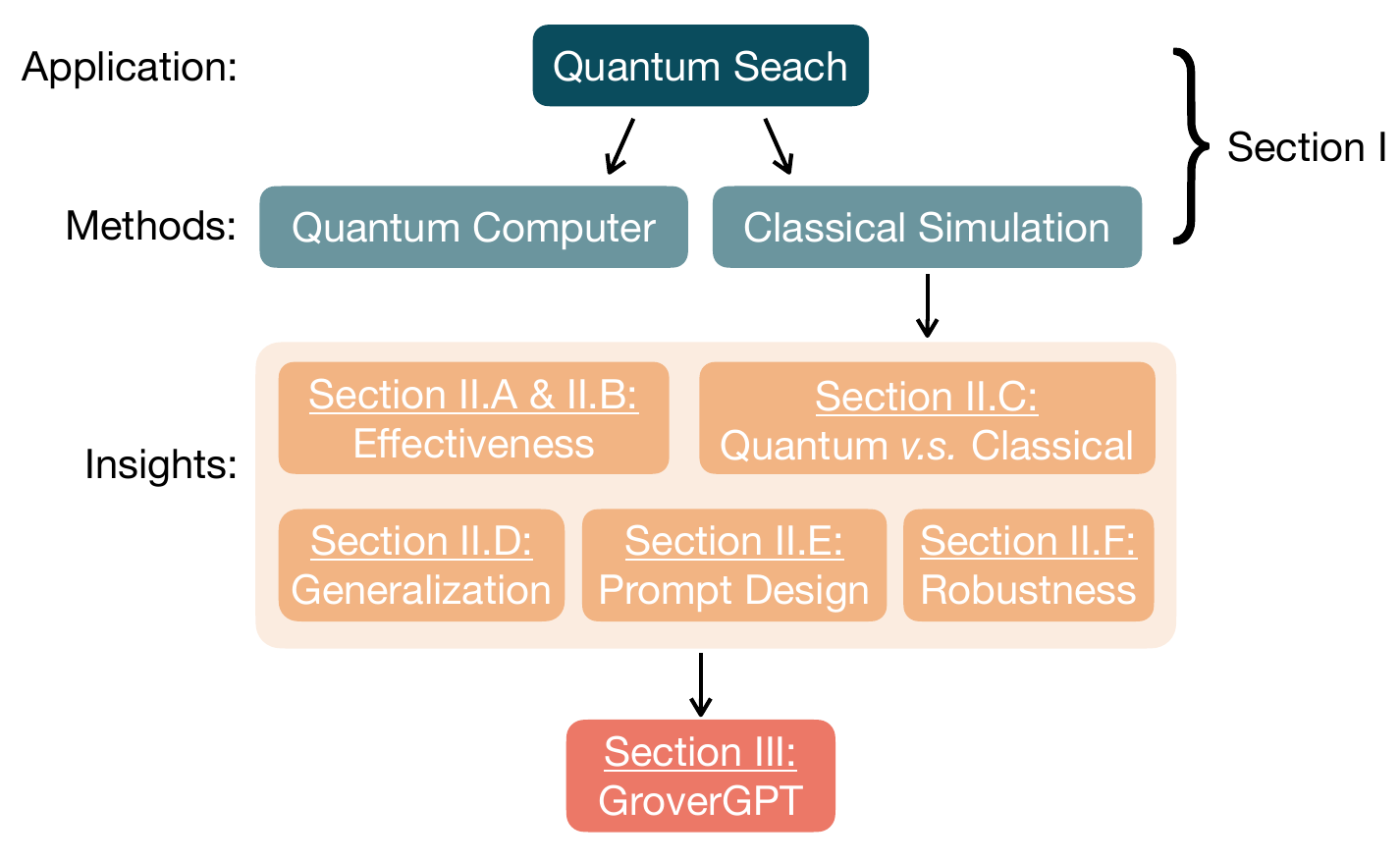}    
    \caption{\textbf{Overview.} We investigate the classical simulation of quantum search through GroverGPT, a large language model approach. Starting from quantum search's implementations via quantum machines and classical simulation, we evaluate GroverGPT along four dimensions: effectiveness of quantum search simulation, generalization from small to large qubit systems, comparative analysis between quantum and classical approaches, and the role of prompt engineering. Through these investigations, GroverGPT demonstrates promising capabilities in bridging quantum-classical computational boundaries.}
    \label{fig:overview}
\end{figure}

In this work, we present an initial investigation into the classical simulation limits of Grover's algorithm using a quantum-focused LLM. \textit{We introduce GroverGPT, the first LLM, to the best of our knowledge, capable of simulating quantum algorithms at an industrial medium scale.} GroverGPT combines quantum circuit simulation with natural language processing (\textit{i.e.}, a branch of AI that enables computers to understand and generate human language) to explore how effectively classical systems can emulate quantum search algorithms. The model is built upon LLaMA's 8-billion architecture,  a state-of-the-art language model developed for processing and generating text trained on a dataset exceeding 15 trillion tokens \cite{vavekanand2024llama}. To simulate a quantum Turing machine, the input of the model will be a classical description of the quantum circuit (in our case it is Grover), and the output will be a sequence of probability distribution for each bit string. 

Our data construction methodology integrates three key components: $1$) \textit{quantum circuit representations} -- diagrams showing sequences of quantum operations, $2$) \textit{quantum assembly language} -- QASM~\cite{cross2022openqasm}, a standardized programming language for describing quantum operations similar to classical computer assembly code, and $3$) natural language interaction. These components are unified through a carefully designed pre-training pipeline based on the Llama architecture~\cite{vavekanand2024llama}. It allows us to begin studying the capability of classical systems to learn and generalize quantum principles without maintaining explicit quantum states.

As a first step toward understanding these simulation capabilities, we analyze our approach using quantum-specific metrics -- \textit{search accuracy} ($\alpha$), \textit{infidelity} ($\epsilon$), and \textit{marked infidelity} ($\epsilon^k$). Our preliminary results show promising performance in simulating quantum search on moderate-sized systems, with encouraging signs of generalization from training on small systems ($3\sim 6$ qubits) to somewhat larger quantum registers (\textit{e.g.}, $20$ qubits). While these initial findings suggest interesting possibilities for classical simulation of quantum algorithms, they also reveal important limitations that appear to be fundamental rather than technical. This exploration provides early insights into the challenges and opportunities at the boundary between classical and quantum computation. This initial investigation contributes to quantum computing research through four key aspects:
\begin{itemize}
    \item A comprehensive dataset comprising $97$K quantum search examples across different qubit sizes ($3\sim 20$ qubits), including quantum circuit simulations, QASM representations, and natural language descriptions, which we release to facilitate further research in quantum algorithm simulation.
    \item A novel experimental framework for studying classical simulation limits of quantum algorithms through language model-based approximation. We released a pre-trained 8-billion-parameter language model (GroverGPT) specialized in quantum search simulation.
    \item Initial empirical observations about how classical systems might learn and generalize quantum principles without explicit quantum state representation. We explore a potential approach to quantum algorithm simulation that aims to balance infidelity with computational efficiency.
\end{itemize}

\begin{figure*}[t]
    \centering
    \includegraphics[width=1.0\textwidth]{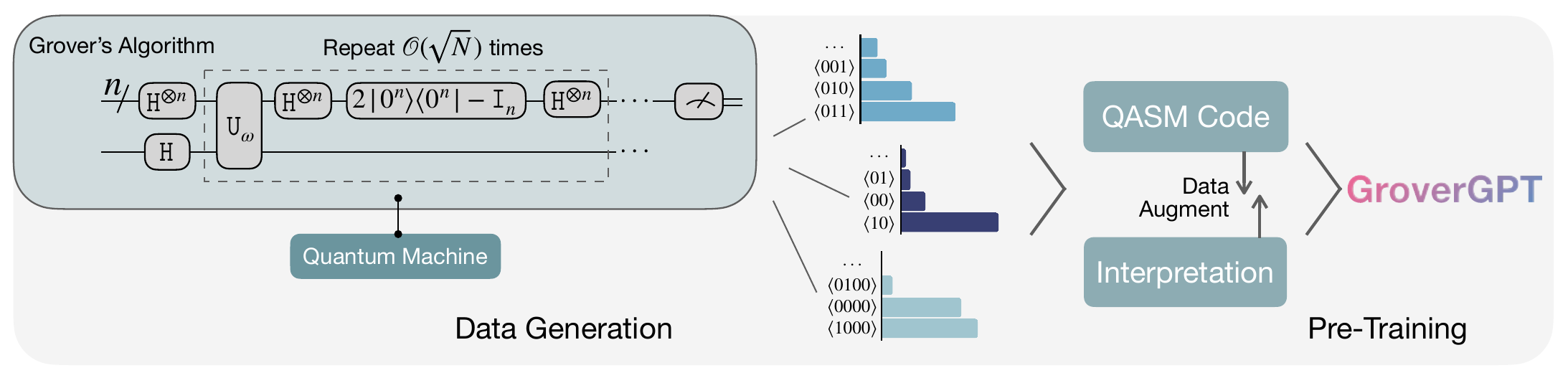}    
    \caption{\textbf{Overview of GroverGPT's pre-training pipeline.} From left to right: \uline{(1)} Data generation begins with implementing Grover's algorithm on a simulated quantum machine, repeating the circuit $\mathcal{O}(\sqrt{N})$ times to construct comprehensive training data where $N=2^n$ and $n$ is the number of qubits. \uline{(2)} The measurement outcomes are collected, represented by probability distributions across different computational basis states (shown in color-coded bars for different qubit configurations). \uline{(3)} The corresponding QASM code is generated to provide standardized circuit descriptions. \uline{(4)} These components are combined through augmented training, integrating both quantum circuit information and measurement data to pre-train the GroverGPT model, which builds upon the Llama-3.1-8B~\cite{vavekanand2024llama} architecture.}
    \label{fig:main}
\end{figure*}

The structure of this paper is organized in Figure \ref{fig:overview}. As a summary, we obtain the following insights:
\begin{itemize}
\item The model can \emph{outperform} general purpose models like OpenAI's GPT-4o (Section \ref{sec:strategy} and \ref{sec:outperform}). 
\item The model can \emph{generalize} towards up to 20 qubits by only looking at few-qubits examples, but the capability drastically decreases with the number of qubits (Section \ref{sec:generalize}). This result indicates that the model somehow knows the structure of the algorithms, but the capability is still limited by the exponential growth of the Hilbert space.
\item The model can learn features of \emph{quantum searching} instead of just learn a \emph{classical searching} algorithm (Section \ref{sec:classical}).
\item The model can learn some structures of quantum circuits with the help of QASM languages as inputs, and some \emph{prompt engineering} frameworks have been developed to make it learn better (Section \ref{sec:prompt}). 
\item The model’s \emph{robustness} is comprehensively evaluated (Section~\ref{sec:robust}), highlighting the impact of training strategies and dataset diversity on performance consistency.
\end{itemize}

Technical methods are summarized in Section \ref{sec:background}, and related knowledge and experimental details have been summarized in Appendix for knowledge completeness. Conclusion and outlook is provided in Section \ref{sec:outlook}.

\section{GroverGPT}

\begin{figure*}
    \centering
    \includegraphics[width=0.85\textwidth]{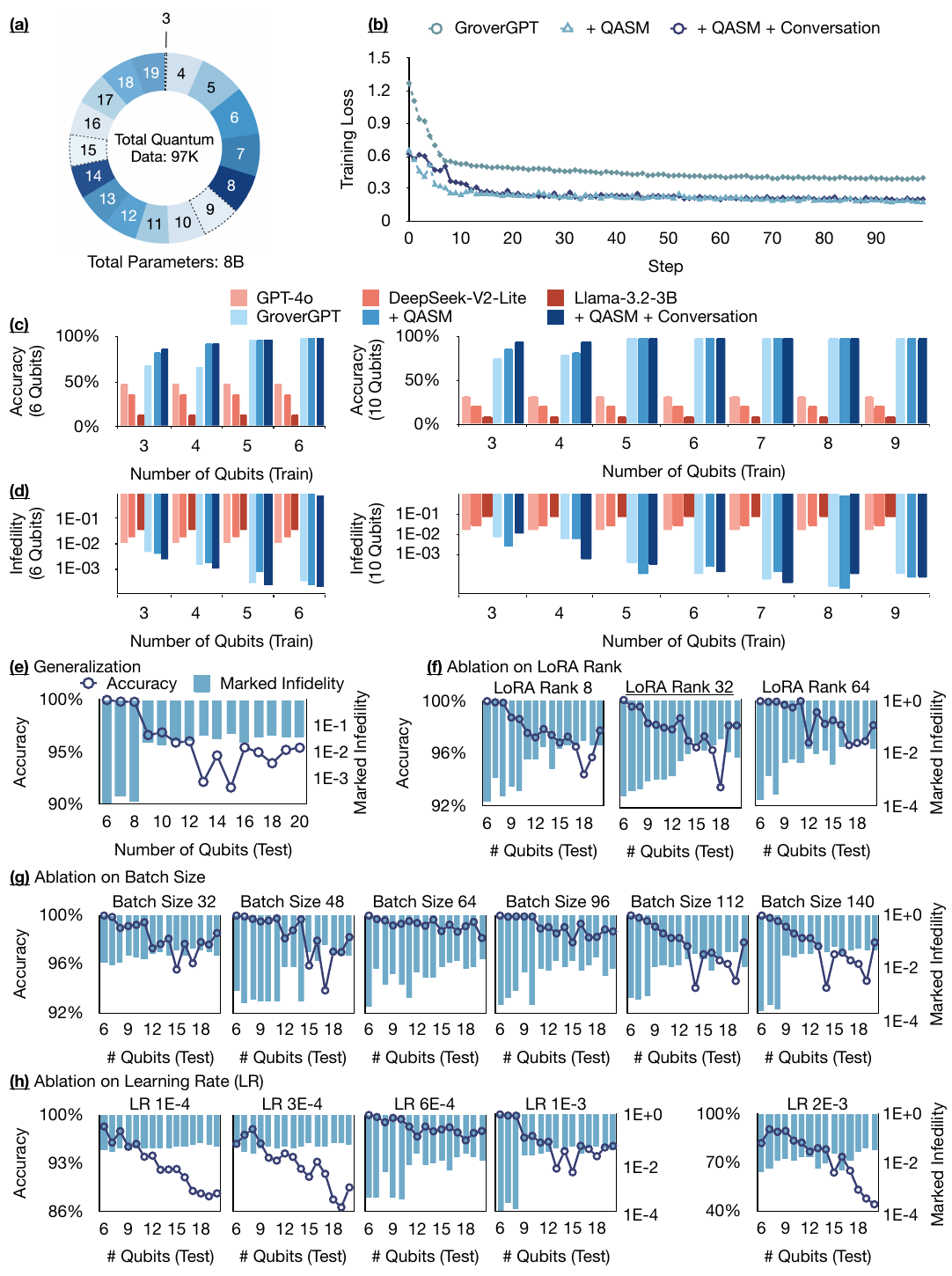}    
    \vspace{-10pt}
    \caption{\textbf{Performance evaluation and generalization capability of GroverGPT.} \textbf{(a)} Distribution of the pre-training dataset comprising $97$K quantum search examples across different qubit sizes from $3$ to $20$. \textbf{(b)} Training loss curves for different GroverGPT variants, showing convergence behavior during pre-training on 3-6 qubit datasets.  \textbf{(c)} Comparative accuracy analysis of GPT-4o, various open-source large models, GroverGPT, and its variants with QASM and conversation components on $6$-qubit (left) and $10$-qubit (right) test sets across different training qubit ranges. \textbf{(d)} \textcolor{black}{Infidelity} $\epsilon$ comparison between models on $6$-qubit (left) and $10$-qubit (right) test sets, demonstrating the error reduction as training qubit count increases. \textbf{(e)} Scalability assessment of GroverGPT trained on $3\sim 6$ qubits, showing accuracy (blue line) and \textcolor{black}{Marked Infidelity} (blue bars) when tested on larger systems ranging from $6$ to $20$ qubits, highlighting the LLM's generalization capabilities beyond its training domain.\textcolor{black}{\textbf{{(f) (g) (h)}} Model performance across a wide range of hyper-parameters (LoRA rank, batch size, learning rate, respectively), highlighting accuracy as a robust indicator.} 
    }
    \label{fig:main-results}
\end{figure*}

\subsection{Pre-Training Strategy Selection for GroverGPT}\label{sec:strategy}
As illustrated in Figure \ref{fig:main}
, we develop GroverGPT through a structured pre-training pipeline built upon the Llama-3.1-8B~\cite{vavekanand2024llama} foundation model. Our training dataset encompasses quantum systems ranging from $3$ to $20$ qubits, carefully separating training ($3\sim 10$ qubits) and testing ($6\sim 20$ qubits) sets. The pre-training process integrates three key components: quantum circuit simulations from Grover's algorithm implementations, corresponding QASM code generated via Qiskit~\cite{qiskit2024}, and natural language conversations about quantum search problems. This multi-modal approach enables GroverGPT to comprehensively understand the structural and behavioral aspects of quantum search operations. Data augmentation through QASM representations and natural language interactions further enhances the model's ability to bridge the gap between abstract quantum algorithms and practical implementations.

\subsection{GroverGPT's Effectiveness of Simulating Quantum Search}\label{sec:outperform}
This section demonstrates GroverGPT's superior performance in quantum search simulation compared to baseline models. Our empirical results demonstrate that GroverGPT significantly outperforms the baseline GPT-4o model in quantum search simulation. As shown in Figure \ref{fig:main-results}c, while GPT-4o maintains a relatively constant accuracy of approximately $45\%$ across different training qubit ranges, GroverGPT achieves substantially higher accuracy, reaching nearly $100\%$ when trained on $5$ or $6$ qubits. This performance gap is particularly evident in both the $6$-qubit (left) and $10$-qubit (right) test scenarios. The integration of QASM and conversation components further enhances GroverGPT's performance, especially in scenarios with fewer training qubits ($3\sim4$ qubits), where the accuracy improves from around $70\%$ to over $80\%$. This consistent improvement pattern suggests that GroverGPT has successfully learned to emulate the fundamental principles of quantum search, rather than merely approximating classical search strategies.

\subsection{Scalability and Generalization Capability of GroverGPT}\label{sec:generalize}
In this section, we examine GroverGPT's remarkable ability to generalize quantum search capabilities from training on small-scale systems to significantly larger quantum systems. GroverGPT exhibits remarkable generalization capabilities across different qubit scales. Figure \ref{fig:main-results}e comprehensively analyzes the model's performance when trained on $3\sim6$ qubits and tested on increasingly larger systems, ranging from $6$ to $20$ qubits. The results show that GroverGPT maintains accuracy above $95\%$ for systems up to $8$ qubits, with only a gradual decline to approximately $95\%$ for systems between $9$ and $20$ qubits. This robust generalization ability is particularly noteworthy given that the model was trained only on smaller systems ($3\sim6$ qubits), suggesting its capacity to extrapolate quantum search principles to substantially larger quantum systems. The parallel trends between accuracy and \textcolor{black}{infidelity} in Figure \ref{fig:main-results}e suggest that the model's performance degradation at larger qubit counts is gradual and predictable, indicating a systematic rather than catastrophic breakdown of simulation capability. This behavior aligns with theoretical expectations about the scalability challenges in quantum simulation and provides valuable insights into the practical limits of classical models in capturing quantum phenomena.

\subsection{Quantum \textit{v.s.} Classical Search Learned by GroverGPT}\label{sec:classical}
Through detailed \textcolor{black}{fidelity analysis} in this section, we investigate whether GroverGPT truly learns quantum search rather than simply approximating classical search strategies. The \textcolor{black}{fidelity analysis} in Figure \ref{fig:main-results}d provides strong evidence that GroverGPT learns genuine quantum search rather than classical approximations. The consistently low infidelity values (below $0.005$ for $6$ qubits and approaching zero for systems trained on $5$ or more qubits) indicate that GroverGPT accurately captures the quantum state amplitudes characteristic of Grover's algorithm. This is in stark contrast to GPT-4o~\cite{openai2024gpt4ocard}'s performance, which shows a consistently high \textcolor{black}{infidelity} (approximately $0.011$), suggesting it fails to capture the subtle quantum features of the search process. Similarly, models such as DeepSeek-V2-Lite \cite{liu2024deepseek,guo2025deepseek} and Llama-3.2-3B~\cite{vavekanand2024llama} exhibit even lower accuracy and higher infidelity values, emphasizing their limitations in capturing the quantum search process. The convergence of infidelity to near-zero values with increased training qubit count is particularly significant, as it indicates that GroverGPT successfully learns the interference patterns and amplitude amplification mechanisms that are quintessential to quantum search. This distinction is further reinforced by the symmetric behavior observed in both $6$-qubit and $10$-qubit test cases, suggesting that the model has internalized fundamental quantum principles rather than memorizing specific problem instances.

\subsection{Impact of Prompt Design Strategies in GroverGPT}\label{sec:prompt}
In this section, we evaluate how different prompting strategies, particularly the integration of QASM code and conversational components, influence GroverGPT's performance in quantum search simulation. The influence of different prompting strategies~\cite{brown2020language,white2023prompt} is clearly illustrated in Figure \ref{fig:main-results}c, where we compare the base GroverGPT model with variants incorporating QASM and conversation components. The addition of QASM prompts improves performance with a substantial margin, particularly evident in the $3\sim4$ qubit training scenarios where accuracy increases by approximately $10\sim15$ percentage points. Further enhancement through conversation components yields additional improvements, particularly notable in the $6$-qubit test case (left panel). This hierarchical improvement pattern suggests that structured quantum circuit descriptions (QASM) combined with natural language interaction create a more robust framework for quantum search simulation. The synergistic effect of these prompting strategies indicates that the model benefits from both the precision of formal quantum circuit descriptions and the contextual richness of natural language interaction. This finding has important implications for the design of future quantum simulation interfaces and educational tools, suggesting that a multi-modal approach to quantum algorithm description might be optimal for both performance and accessibility.

{
\color{black}
\subsection{Robustness of the GroverGPT Training Strategy}\label{sec:robust}

In this section, we evaluate the robustness of GroverGPT by examining the limited variability among its metric indicators and investigating how diverse training sets affect its overall performance.

\textbf{Ablation Study on Training Configurations.}  
Our comprehensive ablation studies investigating LoRA rank, batch size, and learning rate, presented in Figure~\ref{fig:main-results}(f)(g)(h), demonstrate that GroverGPT exhibits remarkable stability across various hyper-parameter configurations. The model maintains consistent performance in terms of both accuracy and infidelity metrics under standard training conditions. However, performance degradation becomes evident at extreme parameter values, particularly when the learning rate falls below $3 \times 10^{-4}$ or exceeds $1 \times 10^{-3}$. These findings underscore GroverGPT's robustness within conventional hyper-parameter ranges, suggesting a practical advantage for real-world implementations.


\begin{figure}
    \centering
    \includegraphics[width=1.01\linewidth]{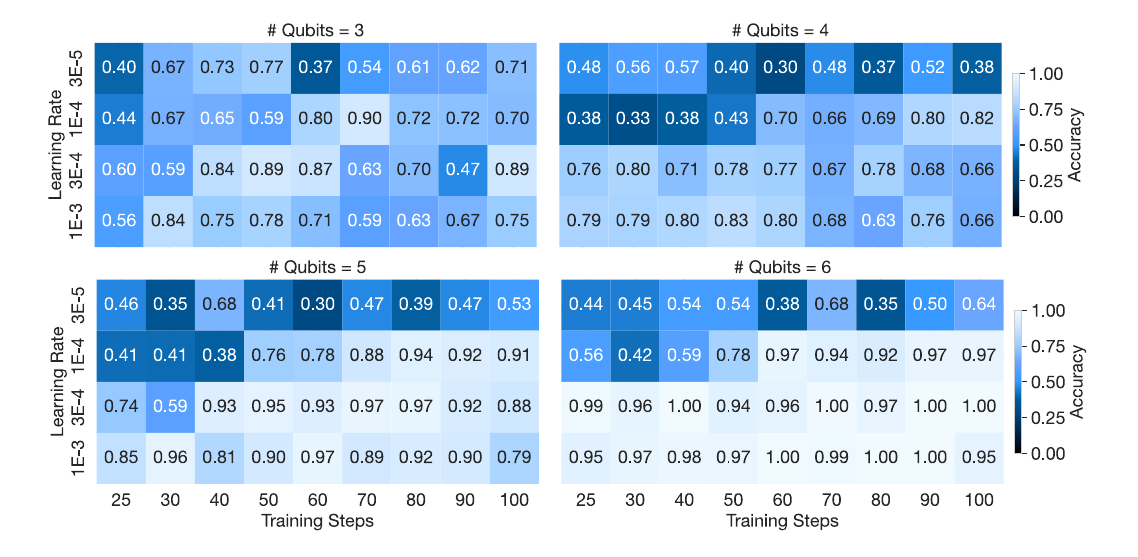}    
    \vspace{-10pt}
    \caption{The model's robustness across diverse training datasets on different qubit systems illustrates its sensitivity to hyper-parameters during finetuning.
    }
    \label{fig:heatmap}
\end{figure}

\textbf{Training Diversity for Robustness.}
Our experimental results demonstrate that training set diversity plays a crucial role in enhancing model robustness. Figure~\ref{fig:heatmap} illustrates this through four heatmaps generated under different hyper-parameter configurations. We systematically expanded the training dataset from a single qubit configuration (\textit{e.g.}, qubits=$3$) to increasingly diverse combinations (\textit{e.g.}, qubits=$[3,4]$, $[3,4,5]$, and $[3,4,5,6]$). The results reveal that as training data diversity increases, regions of high accuracy become more distinctly defined within the hyper-parameter space, facilitating more effective optimization. These findings suggest that enhanced data diversity not only strengthens model robustness but may also improve generalization capabilities across other challenging scenarios.
}

\section{Methodology}\label{sec:background}
\subsection{Grover's Algorithm}
As demonstrated in~\cite{grassl2016applying,grover1996fast,mavroeidis2018impact}, Grover's Algorithm~\cite{grover1996fast} represents a fundamental breakthrough in quantum computing, demonstrating significant computational advantages over classical methods. For the ubiquitous task of searching through an unstructured database of $N$ items, classical computers require examining items one by one, taking $O(N)$ operations on average. In contrast, Grover's Algorithm achieves a quadratic speedup, requiring only $O(\sqrt{N})$ operations by leveraging principles of quantum mechanics.

As indicated by Figure\ref{fig:main-results} (upper left), the algorithm creates a quantum state that simultaneously represents all possible database items through superposition -- a quantum property where a qubit can exist in multiple states simultaneously. It then iteratively applies two operations: one that marks the desired solution and another that amplifies the probability of finding this marked state. After $\sim\sqrt{N}$ iterations, measuring the system yields the target item with high probability.

\subsection{Classical Simulation of GroverGPT}

We adopt a noise-free classical simulation of Grover's quantum search algorithm using Qiskit~\cite{qiskit2024}, which is an open-source software development kit for quantum computing developed by IBM, and its state vector simulator to establish ground truth data for training and evaluation. This simulator tracks the complete mathematical description of a quantum system: for an $n$-qubit system where a qubit is the quantum equivalent of a classical bit, the simulator maintains a state vector of dimension $N=2^n$. It enables the exact computation of quantum state amplitudes, which are the complex numbers that describe the probability of measuring each possible quantum state, along with their measurement probabilities.

As shown in Figure\ref{fig:main}
, the simulation executes the standard Grover circuit template, which is a sequence of quantum operations, with $\mathcal{O}(\sqrt{2^n})$ iterations. This process comprises three essential components: initialization through Hadamard gates, which are quantum operations that create equal superpositions; oracle operations for target state marking, which are operations that identify the solution we're searching for; and diffusion operators for amplitude amplification, which are operations that enhance the probability of finding the marked state. Through multiple simulation shots, which are repeated runs of the quantum circuit, we obtain relatively precise probability distributions over computational basis states, which represents the possible measurement outcomes. This provides reliable reference data for assessing GroverGPT's quantum search simulation capabilities.


\subsection{Evaluation Metrics}
To rigorously assess GroverGPT's quantum search simulation capabilities, we establish three complementary evaluation metrics. Given a quantum system with $n$ qubits and $k$ marked states, let $|\psi_{\text{final}}\rangle$ represent the final quantum state after applying Grover's algorithm. The measurement outcome probability distribution is defined as:
\begin{equation}
\mathcal{P} = {(s_i, p_i) \mid i \in [2^n], p_i = |\langle s_i|\psi_{\text{final}}\rangle|^2}
\end{equation}
where $s_i$ represents the $i$-th computational basis state and $p_i$ its corresponding measurement probability. Let $\mathcal{P}_{\text{model}}$ and $\mathcal{P}_{\text{true}}$ denote the probability distributions generated by GroverGPT and the ideal quantum simulator, respectively. We evaluate performance using:
\begin{itemize}
\item \textbf{Search Accuracy ($\alpha$)}: Measures the model's ability to identify marked states correctly, defined as:
\begin{equation}
\alpha = \frac{|\mathcal{M}_{\text{model}} \cap \mathcal{M}_{\text{true}}|}{k}
\end{equation}
where $\mathcal{M}_{\text{model}}$ and $\mathcal{M}_{\text{true}}$ are the sets of $k$ highest-probability states in $\mathcal{P}_{\text{model}}$ and $\mathcal{P}_{\text{true}}$, respectively.
\item \textbf{Infidelity~($\epsilon$)}: Quantifies the overall quantum state reproduction accuracy through:
\begin{equation}
\epsilon = \frac{1}{2^n}\sum_{i=1}^{2^n} (p_i^{\text{model}} - p_i^{\text{true}})^2
\end{equation}
where $p_i^{\text{model}}$ and $p_i^{\text{true}}$ are probabilities from $\mathcal{P}_{\text{model}}$ and $\mathcal{P}_{\text{true}}$.
\item \textbf{Marked Infidelity~($\epsilon^k$)}: Specifically evaluates the accuracy of marked state probability predictions:
\begin{equation}
\epsilon^k = \frac{1}{k}\sum_{s_i \in \mathcal{M}_{\text{true}}} (p_i^{\text{model}} - p_i^{\text{true}})^2
\end{equation}
\end{itemize}
These metrics provide complementary perspectives on GroverGPT's performance: $\alpha$ assesses the model's ability to identify correct search solutions, $\epsilon$ evaluates the overall quantum state reproduction infidelity, and $\epsilon^k$ focuses specifically on the accuracy of marked state predictions. Together, they comprehensively evaluate the model's practical search capabilities and theoretical quantum properties.

\section{Conclusion and Outlook}\label{sec:outlook}
In this work, we demonstrate that an industrial-level, medium-scale LLM can be designed to simulate noiseless quantum algorithms, such as Grover's search, and potentially outperform popular general-purpose models like OpenAI's GPT-4. This is achieved through the development of GroverGPT. Additionally, we present evidence suggesting that the model can learn and generalize certain features of quantum algorithms, though its performance significantly deteriorates as the number of qubits increases. Our work offers a valuable tool for advancing research and education in quantum algorithms.

Our work opens a new avenue for exploring the boundaries of classical simulability and quantum advantage using LLMs. This raises numerous questions for future research. For example, how well can LLMs simulate noisy quantum computers? Could they effectively model current noisy quantum systems with 100 to 1,000 qubits, and what level of error would be tolerable? Can other quantum algorithms be simulated, or might it even be possible to develop a foundational model capable of simulating quantum Turing machines? Can we simulate quantum error correction codes? Furthermore, if resources from leading LLM developers such as OpenAI, Anthropic, or xAI were available for training models specifically tailored to quantum algorithms, how many qubits and what circuit depth could we feasibly simulate? These intriguing questions remain open for future investigation.

\section*{Acknowledgment}
We thank Yangrui Hu, Jin-Peng Liu, Ziwen Liu, Zhiding Liang and Suqiong Zeng for discussion and help. MC, JC, and JL are supported in part by the University of Pittsburgh, School of Computing and Information, Department of Computer Science, Pitt Cyber, and the PQI Community Collaboration Awards. PL and TC are supported in part by Cisco Faculty Award and UNC SDSS Seed Grant.

\bibliographystyle{naturemag}

\section*{Appendix}

\subsection{Preliminaries for Quantum Computing}

\textbf{Quantum Turing Machine.} A \textit{quantum Turing machine} is a common computational model we use for universal quantum computing \cite{nielsen2010quantum}. Here is a simplified and informal definition of a quantum Turing machine. For an $n$-qubit quantum Turing machine, we define a linear space $\mathcal{H}$, where we call it the Hilbert space (a complex linear space equipped with an inner product), such that $\dim \mathcal{H} =2^n$, whose basis vector is the bit string denoted as $\ket{x_0,x_1,\cdots,x_{n-1}}$ where $x_i\in \{0,1\}$ for $0\le i\le n-1$. Thus, the input of the machine is a description of a sequence $\mathcal{U}=\{U_1,U_2,\cdots,U_L\}$ of unitary matrices on $\mathcal{H}$. We call $\mathcal{U}$ the \textit{quantum circuit}, and we call $U_a$ for $1\le a\le L$ \textit{quantum gates}. The output of the model is a probability distribution on the bit string $\ket{x_0,x_1,\cdots,x_{n-1}}$ where $x_i\in \{0,1\}$ for $0\le i\le n-1$. The probability distribution $p_i$ is determined by the \textit{Born rule}, $p_i=\abs{\bra{x_i}U_1U_2\cdots U_L\ket{00,\cdots,0}}^2$. In practice, what we receive from quantum computers are samples of the bit string exactly following the distribution $p_i$. Due to the central limit theorem, when the number of samples is large, it is not hard to approximate the original probability distribution. Sometimes, we \textit{prepare a different initial state} by changing $\ket{00,\cdots,0}$ to something else in the formula of $p_i$. However, it is equivalent to redefining $\mathcal{U}$. For a more detailed introduction of all those ingredients, see the following paragraphs for a more detailed illustration. 

\textbf{Quantum Computing and Its Key Principles.} Quantum computing is operated on quantum computers or quantum devices. It leverages quantum mechanics to conduct different computing tasks. In some specific applications including encryption~\cite{shor1994algorithms}, quantum simulation~\cite{kim2023evidence} and quantum machine learning~\cite{biamonte2017quantum,ajagekar2020quantum,ajagekar2021quantum,liu2024towards} etc., with a large-scale quantum computer, quantum computing can theoretically process exponentially faster speed than current classical computing.

One of the differences between quantum computing and classical computing lies on the basic computing unit. The classical \textit{Bit}, which is the fundamental concept of classical computing, is either at state 0 or 1, while the \textit{Quantum Bit}, or \textit{qubit}, which is the fundamental concept of quantum computing, could stay in \(|0\rangle\) or \(|1\rangle\), or "between" these two \textit{computational basis states}. It is termed the \textit{superposition}:

\begin{equation}
    |\psi\rangle = \alpha |0\rangle + \beta |1\rangle ,
\end{equation}
where \(\alpha\) and \(\beta\) are the corresponding amplitudes for each basis state.

Basically, quantum computing will result in the change of a single qubit or multiple qubits. This process can also be described as \textit{quantum information processing}. After \textit{measurement}, each qubit can only output a single bit of information. Measurement of a qubit means changing the state of every single qubit by collapsing it from its superposition of  \(|0\rangle\) and \(|1\rangle\) to either \(|0\rangle\) or \(|1\rangle\) state depending on the probabilities. Measurement is one way that causes \textit{decoherence}, which refers to the process which a quantum state collapses into a non-quantum state. Besides, quantum systems follow a key principle termed \textit{entanglement}, which refers to the phenomenon that a qubit possess the ability to correlate its state with other qubits. Meanwhile, superposition and entanglement offer the condition for \textit{interference}, which refers to the phenomenon that entangled qubits, each with multiple states, can interfere with each other, leading to amplifying or discouraging the probabilities, denoted as constructive interference and destructive interference respectively.

\textbf{Quantum Gates and Quantum Circuits.} Quantum computing relies on operations in quantum circuits, which contain reversibly elementary \textit{quantum logic gates}, or so-called \textit{quantum gates}. Quantum gates can be also represented as unitary operators. A unitary operator or matrix \(U\) on a Hilbert Space \(\mathcal{H}\) indicates the following fact:

\begin{equation}
    U^\dagger U = I,
\end{equation}
where \(I\) is the identity matrix and \(U^\dagger
\) is the adjoint or complex conjugate of the matrix \(U\).

We leverage several frequently adopted quantum gates to construct the quantum circuit for implementing the Grover's quantum search algorithm, including the Hadamard gate or so-called \(H\) gate which turns a $|0\rangle$ into $(|0\rangle + |1\rangle)/\sqrt{2}$ and turns $|1\rangle$ into $(|0\rangle - |1\rangle)/\sqrt{2}$, single-qubit quantum \textit{NOT} gate or so-called \(X\) gate, \(Z\) gate which leaves \(|0\rangle\) and flips the sign of \(|1\rangle\) or \(-|1\rangle\): 

\begin{equation}
\begin{aligned}
    &H \equiv \frac{1}{\sqrt{2}} \left[ \begin{array}{cc} 1 & 1 \\ 1 & -1 \end{array} \right], X \equiv \left[ \begin{array}{cc} 0 & 1 \\ 1 & 0 \end{array} \right], Z \equiv \left[ \begin{array}{cc} 1 & 0 \\ 0 & -1 \end{array} \right]
\end{aligned}
\end{equation}

\begin{figure}[H] 
    \centering
    \includegraphics[width=0.4\textwidth]{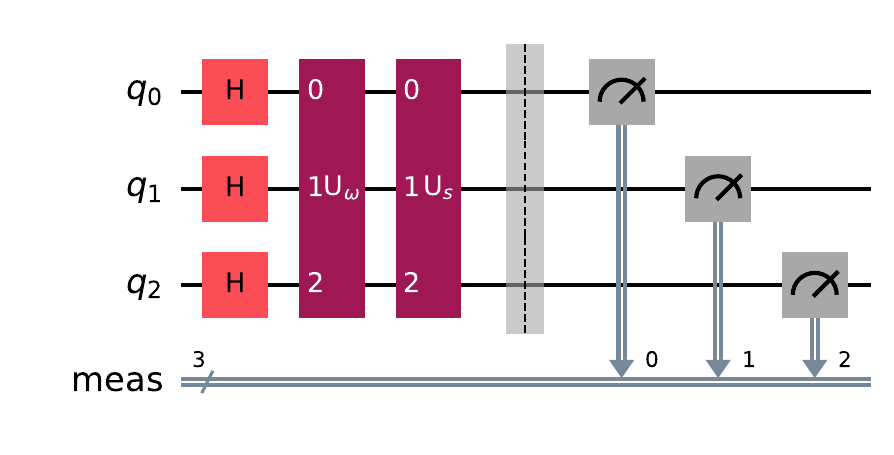}    
    \caption{Plotted circuit of Grover's searching algorithm implemented using Qiskit under 3 qubits.}
    \label{Grover_Circuit_qiskit}
\end{figure}

Quantum circuits are models for quantum computing and quantum algorithms. Basic components include quantum gates, measurements, and possibly some other actions. Fig.\ref{Grover_Circuit_qiskit} gives an example of what a quantum circuit might look like, specifically by showing the plotted circuit of Grover's searching algorithm using Qiskit.

\textbf{Open Quantum Assembly Language~(OpenQASM).} In this study, we leverage the \textit{OpenQASM} \cite{cross2017open} programming language to describe the quantum circuits and Grover's searching algorithm. OpenQASM is designed to serve as an intermediate representation to allow communication between the high-level compilers and the quantum hardware. It is implemented using \textit{python}.
By default, the version of the OpenQASM we adopt is 
\textbf{Version 3.0}. Below we provide an example when defining a quantum circuit that implements  Grover's quantum searching algorithm under 3 qubits:

\begin{lstlisting}[caption=OpenQASM description for the Grover's searching algorithm under 3 qubits.]
OPENQASM 3.0;
include "stdgates.inc";
gate U$_\omega$ _gate_q_0, _gate_q_1, _gate_q_2 {
  cz _gate_q_0, _gate_q_2;
  cz _gate_q_1, _gate_q_2;
}
gate U$_s$ _gate_q_0, _gate_q_1, _gate_q_2 {
  h _gate_q_0;
  h _gate_q_1;
  h _gate_q_2;
  x _gate_q_0;
  x _gate_q_1;
  x _gate_q_2;
  h _gate_q_2;
  ccx _gate_q_0, _gate_q_1, _gate_q_2;
  h _gate_q_2;
  x _gate_q_0;
  x _gate_q_1;
  x _gate_q_2;
  h _gate_q_0;
  h _gate_q_1;
  h _gate_q_2;
}
bit[3] meas;
qubit[3] q;
h q[0];
h q[1];
h q[2];
U$_\omega$ q[0], q[1], q[2];
U$_s$ q[0], q[1], q[2];
barrier q[0], q[1], q[2];
meas[0] = measure q[0];
meas[1] = measure q[1];
meas[2] = measure q[2];
\end{lstlisting}

\subsection{Preliminaries for Grover's Algorithm}

Grover's algorithm, introduced by Lov Grover~\cite{grover1996fast}, is a quantum algorithm designed to search for a specific item within an unsorted database of $N$ items. In contrast to classical algorithms, which require an average of $O(N)$ operations to find the target item, Grover's algorithm achieves this task in $O(\sqrt{N})$ steps, offering a quadratic speedup. This makes it a powerful tool for various applications, including database search, optimization problems~\cite{chakrabarty2017dynamic}, and cryptographic analysis~\cite{sarah2024practical}.

The algorithm consists of several key components: the initial state preparation, the oracle, the diffusion operator, and the measurement. Each of these components plays a crucial role in the algorithm's ability to amplify the amplitude of the target state, which increases the probability of finding it.

\textbf{Initial State Preparation.}  The algorithm begins by preparing an initial state that is a uniform superposition of all possible states in the $N$-dimensional Hilbert space. If the $N$ items are indexed by $n$-qubit states, the initial state is:

$$
|\psi_0\rangle = \frac{1}{\sqrt{N}} \sum_{x=0}^{N-1} |x\rangle
$$

This state is prepared by applying Hadamard gates ($H$) to all $n$ qubits initialized in the $|0\rangle$ state:

$$
H^{\otimes n} |0\rangle = \frac{1}{\sqrt{2^n}} \sum_{x=0}^{2^n-1} |x\rangle
$$

\textbf{The Oracle.}The oracle is a quantum subroutine that marks the target state(s) by flipping their phase. For a single marked state $|x_t\rangle$, the oracle operator $O$ is defined as:

$$
O = I - 2|x_t\rangle\langle x_t|
$$

This operator applies a phase flip to the target state:

$$
O|x\rangle = 
\begin{cases} 
-|x\rangle & \text{if } x = x_t \\
|x\rangle & \text{otherwise}
\end{cases}
$$

The implementation of the oracle depends on the specific problem being solved. In general, it involves encoding the target state $x_t$ using a unitary circuit that compares the input state to $x_t$ and applies a phase shift (e.g., a $Z$-gate) conditioned on the comparison result. For example, if $x_t$ is known, the oracle can be implemented using a controlled-$Z$ gate, where the control qubits are those that specify $x_t$.

\textbf{The Diffusion Operator.}The diffusion operator, also known as the Grover operator, is responsible for amplifying the amplitude of the target state. It is defined as:

$$
D = 2|\psi_0\rangle\langle\psi_0| - I
$$

where $|\psi_0\rangle$ is the initial uniform superposition state and $I$ is the identity operator. The diffusion operator can be implemented using a sequence of Hadamard gates, a multi-qubit $Z$-gate, and another sequence of Hadamard gates. Specifically, the diffusion operator can be written as:

$$
D = H^{\otimes n} (2|0\rangle\langle 0| - I) H^{\otimes n}
$$

This operator effectively inverts the amplitudes of the quantum state to the average amplitude.
The average amplitude $ \langle \alpha \rangle $ before diffusion is:
$$
\langle \alpha \rangle = \frac{N-2}{N\sqrt{N}}
$$
The diffusion operator transforms each amplitude $ \alpha_x $ to:
$$
\beta_x = 2\langle \alpha \rangle - \alpha_x
$$
For the target state $ |x_t\rangle $:
$$
\beta_{x_t} = 2\langle \alpha \rangle - (-\frac{1}{\sqrt{N}}) = \frac{3N - 4}{N\sqrt{N}}
$$
For other states $ |x\rangle $ (where $ x \neq x_t $):
$$
\beta_x = 2\langle \alpha \rangle - \frac{1}{\sqrt{N}} = \frac{N - 4}{N\sqrt{N}}
$$
Repeating the oracle and diffusion steps iteratively increases the amplitude of the target state. In the Hilbert space, the diffusion operator reflects the state vector about the average amplitude vector, which is constructive interference for the target state.

\textbf{Complexity of Grover's Algorithm. }A single Grover iteration consists of applying the oracle $ O $ followed by the diffusion operator $ D $. If the initial state is $ |\psi_0\rangle $, the state after $ k $ iterations is:

$$
|\psi_k\rangle = (D \cdot O)^k |\psi_0\rangle
$$

The optimal number of iterations $k$ to maximize the probability of measuring a marked state is approximately:

\begin{equation*}
k \approx \frac{\pi}{4} \sqrt{\frac{N}{M}},
\end{equation*}
where $M$ is the number of marked states. This formula generalizes the scenario for multiple marked states, reducing to $k \approx \frac{\pi}{4} \sqrt{N}$ when $M=1$.

The amplitudes of the marked and unmarked states evolve with each iteration. Specifically, the amplitude of the marked states is given by:

\begin{equation*}
a_t^{(k)} = \sin\left((2k+1)\theta\right),
\end{equation*}
where $\theta = \arcsin\left(\sqrt{\frac{M}{N}}\right)$. 

This evolution indicates that each iteration amplifies the amplitude of the marked states. The optimal $k$ is chosen such that $(2k+1)\theta \approx \frac{\pi}{2}$, ensuring the probability of measuring a marked state is maximized.
Grover's algorithm thus achieves its quadratic speedup by iteratively increasing the amplitude of the marked states with the power of quantum interference.

\subsection{Preliminaries for Large Language Models}
\textbf{Language Models and Their Development.}
Language models (LMs) serves as a key method for enhancing machine language understanding, At its core, LMs maximize the probabilistic likelihood structure of word sequences, allowing for predictions of upcoming or missing words. This foundational capability supports a wide range of natural language processing (NLP) applications, including tasks like machine translation and conversational systems.

The recent success of pre-trained language models (PLMs) has demonstrated that increasing model size, training data volume, or computational resources often enhances their ability to perform downstream tasks. This observation, commonly referred to as the scaling law, has driven the development of large-scale models. Models like GPT and PaLM mark a major advancement, showcasing the ability to solve complex tasks and generalize from limited examples, highlighting the critical role of scaling in enhancing model performance.

Building on the advancements in LLMs, Meta introduced the Llama (Large Language Model Meta AI) series, which features open-source LLMs optimized for performance and accessibility.

\textbf{LLMs basic Architecture. }The Llama models are built upon the Transformer architecture, which is renowned for their self-attention mechanism and modular design. While standard Transformer models consist of both encoder and decoder stacks, Llama focuses exclusively on the Transformer decoder. 

\begin{itemize}
    \item \textbf{Transformer architecture:}
    Each Transformer decoder layer comprises a multi-head attention mechanism to capture dependencies within the sequence being generated, a feed-forward network (FFN) to enhance model expressivity, and residual connections with normalization to improve training stability. 
    
    \item \textbf{Self-Attention Mechanism:} The self-attention mechanism allows the model to capture dependencies between tokens in the input sequence. It operates by mapping a query (\(Q\)), key (\(K\)), and value (\(V\)) to an output, computed as:
    \begin{equation*}    
    \text{Attention}(Q, K, V) = \text{softmax}\left(\frac{Q K^\top}{\sqrt{d_k}}\right) V,
    \end{equation*}
    where \(d_k\) is the dimensionality of the keys. The query, key, and value tensors are derived from the input sequence using learned linear transformations.
    \item \textbf{Feed-Forward Network (FFN):}  
    The FFN in the Transformer decoder enhances the model's ability to represent complex patterns through independent non-linear transformations at each sequence position. The FFN is expressed as:
    \[
    \text{FFN}(x) = \text{ReLU}(xW_1 + b_1)W_2 + b_2,
    \]
    where \(x\) is the input, \(W_1\) and \(W_2\) are weight matrices, and \(b_1\) and \(b_2\) are biases. 

    \item\textbf{Layer Normalization (LayerNorm):}  
    LayerNorm is applied within each decoder layer to stabilize and accelerate training by normalizing the input to each sub-layer. For an input \(x\), the normalized output is computed as:
    \[
    \text{LayerNorm}(x) = \frac{x - \mu}{\sigma} \cdot \gamma + \beta,
    \]
    where \(\mu\) is the mean,  \(\sigma\) is the standard deviation, \(\gamma, \beta\) are learnable scaling and shifting parameters. This component ensures that the input to each sub-layer remains well-scaled, which helps mitigate exploding or vanishing gradients in deep networks.
    
\end{itemize}

\textbf{Key Improvements of Llama Models.}
The Llama models introduce several enhancements to the Transformer decoder to optimize both computational efficiency and expressivity for text generation tasks:
\begin{itemize}
    \item \textbf{Grouped Query Attention (GQA):}
    To improve the efficiency of the self-attention mechanism, Llama employs GQA, which groups multiple query heads to share the same key-value projections. In GQA, the attention computation is modified as:
    \begin{equation*}  
    \text{Attention}(Q_g, K_g, V_g) = \text{softmax}\left(\frac{Q_g K_g^\top}{\sqrt{d_k}}\right) V_g,
    \end{equation*}
    where \(Q_g\), \(K_g\), and \(V_g\) are grouped projections. GQA allows for fewer key-value caches during inference and speeds up the decoding process.

    \item \textbf{Root Mean Square Normalization (RMSNorm):}
    Llama employs RMSNorm instead of applying layer normalization to the output. RMSNorm computes the normalized vector as:
    \begin{equation*} 
    \text{RMSNorm}(x) = \frac{x}{\text{RMS}(x)} \cdot \gamma,
    \end{equation*}
    where
    \begin{equation*} 
    \text{RMS}(x) = \sqrt{\frac{1}{d} \sum_{i=1}^d x_i^2}, \quad x \in \mathbb{R}^d
    \end{equation*}
    \(x\) is the input vector, \(d\) is its dimensionality, and \(\gamma \in \mathbb{R}^d\) is a learnable scaling parameter. Unlike LayerNorm, RMSNorm does not include a bias term, which reduces computational overhead.

    \item \textbf{SwiGLU Activation Function:}
    Llama replaces the conventional ReLU activation function with SwiGLU to achieve a balance between computational efficiency and expressivity, which is defined as:
    \begin{equation*} 
    \text{SwiGLU}(x) = \text{GELU}(xW_1 + b_1) \odot (xW_2 + b_2)
    \end{equation*}
    where \(W_1, W_2 \in \mathbb{R}^{d_{\text{in}} \times d_{\text{out}}}\) are weight matrices, \(b_1, b_2 \in \mathbb{R}^{d_{\text{out}}}\) are biases, and \(\odot\) denotes element-wise multiplication. The Gaussian Error Linear Unit (GELU) is computed as:
    \begin{equation*} 
    \text{GELU}(z) = z \cdot \Phi(z), \quad \Phi(z) = \frac{1}{2}\left[1 + \text{erf}\left(\frac{z}{\sqrt{2}}\right)\right]
    \end{equation*}

\end{itemize}

\subsection{Llama Pre-Training Details for Initializing GroverGPT}
\textbf{Pre-Training Datasets. }The pre-training dataset for Llama 3 is curated from diverse sources containing knowledge up to 2023, with strict removal of PII (Personally Identifiable Information) and adult content. Web data is cleaned using custom parsers, retaining structure for math and code, and applying URL, document, and line-level de-duplication. Heuristic filters and model-based classifiers (e.g., DistilRoberta) ensure high-quality tokens by removing low-quality and repetitive content. Specialized pipelines extract math, reasoning, and code data with prompt-tuned models for STEM-specific tasks. Multilingual data is processed with FastText for language classification (176 languages) and quality-ranked using a multilingual Llama 2-based classifier. The final data mix includes 50\% general knowledge, 25\% math and reasoning, 17\% code, and 8\% multilingual data. Annealing on 40M tokens improves performance, with a 24.0\% gain on GSM8k and 6.4\% on MATH for the 8B model. The total token counts used in pre-training is around 15T+. Scaling law experiments guide the optimal data mix for high downstream task performance.

\textbf{Pre-Training Process. }The pre-training recipe for Llama is carefully designed to ensure model stability and maximize performance across diverse tasks. The pre-training process is divided into three distinct stages: {initial pre-training}, {long-context pre-training}, and {annealing}. Each stage is described below:

\begin{itemize}
    \item \textbf{Initial pre-training. }The initial phase of training uses the AdamW optimizer with a peak learning rate of $8 \times 10^{-5}$, following an $8,000$-step linear warm-up and cosine decay over $1,200,000$ steps. Batch size and sequence length start at $4$M tokens and $4,096$ tokens, doubling to $8$M and $8,192$ tokens after $252$M tokens and again to $16$M after $2.87$T tokens. The training data mix is dynamically adjusted by increasing non-English data, upsampling mathematical datasets, adding recent web data, and downsampling low-quality subsets to enhance multilingual and task-specific performance.
    \item \textbf{Long-context pre-training. }To enable Llama to process long contexts of up to $128$K tokens, the context length is gradually increased from $8$K to $128$K in six stages, using approximately $800$B tokens. Successful adaptation is assessed by recovering short-context performance and solving "needle in a haystack" tasks for long sequences.
    \item \textbf{Annealing. }The final stage of pre-training anneals the learning rate to $0$ over the last $40$M tokens, upsampling high-quality data sources and applying Polyak averaging to produce the final model.
\end{itemize}

\subsection{GroverGPT Fine-Tuning Details}
\textbf{Loss Function.} During the supervised fine-tuning (SFT) phase, the LLM is optimized using a standard cross-entropy loss to align its predictions with target outputs. The loss function is defined as:
\[
\mathcal{L}_{\text{SFT}} = - \frac{1}{N} \sum_{i=1}^{N} \sum_{t=1}^{T_i} \log P_\theta(y_{i,t} | y_{i,<t}, x_i)
\]
where \(N\) is the number of training samples, \(T_i\) is the length of the target sequence for the \(i\)-th sample, \(y_{i,t}\) is the ground truth token, \(y_{i,<t}\) represents the preceding tokens, \(x_i\) is the input prompt, and \(P_\theta(y_{i,t} | y_{i,<t}, x_i)\) is the model’s predicted probability. This loss function ensures that the model learns to predict target tokens accurately based on the provided context and previously predicted tokens.
\newpage
\textbf{GroverGPT Prompt. }We use the following prompt to finetune the Llama models to simulate Grover's algorithm. The prompt is provided to the Llama-3.1-8B model to generate the desired responses. The simplest version of the prompt does not include QASM instructions. Below is an example of the prompt with QASM and its corresponding question-answer pair:

\begin{tcolorbox}[colframe=black!95, colback=white!115, sharp corners]
\textbf{Prompt:}  
\\
Question:  
\\I want you to act as a quantum computer specialized in performing Grover's algorithm.  
I will type a circuit, and you will reply with what a quantum computer should output.  
I want you to only reply with the output in a dictionary that contains the top-30 probabilities and nothing else. The input marked status is: 0000 for a 4-qubit system.  
\\Here is the QASM circuit: \\"h q[0]; h q[1]; h q[2]; h q[3]; x q[0]; x q[1]; x q[2]; x q[3]; h q[3]; mcx\_0 q[0], q[1], q[2], q[3]; h q[3]; x q[0]; x q[1]; x q[2]; x q[3]; h q[0]; h q[1]; h q[2]; h q[3]; x q[0]; x q[1]; x q[2]; x q[3]; h q[3]; mcx\_1 q[0], q[1], q[2], q[3]; h q[3]; x q[0]; x q[1]; x q[2]; x q[3]; h q[0]; h q[1]; h q[2]; h q[3]; x q[0]; x q[1]; x q[2]; x q[3]; h q[3]; mcx\_2 q[0], q[1], q[2], q[3]; h q[3]; x q[0]; x q[1]; x q[2]; x q[3]; h q[0]; h q[1]; h q[2]; h q[3]; x q[0]; x q[1]; x q[2]; x q[3]; h q[3]; mcx\_3 q[0], q[1], q[2], q[3]; h q[3]; x q[0]; x q[1]; x q[2]; x q[3]; h q[0]; h q[1]; h q[2]; h q[3]; x q[0]; x q[1]; x q[2]; x q[3]; h q[3]; mcx\_4 q[0], q[1], q[2], q[3]; h q[3]; x q[0]; x q[1]; x q[2]; x q[3]; h q[0]; h q[1]; h q[2]; h q[3]; x q[0]; x q[1]; x q[2]; x q[3]; h q[3]; mcx\_5 q[0], q[1], q[2], q[3]; h q[3]; x q[0]; x q[1]; x q[2]; x q[3]; h q[0]; h q[1]; h q[2]; h q[3];"

Answer: \\ 
\{'0000': 0.9596, '0001': 0.0037, '0101': 0.0032, '0100': 0.0032, '0110': 0.0031, '0111': 0.003, '0010': 0.003, '1000': 0.0028, '1100': 0.0028, '1110': 0.0027\}
\end{tcolorbox} 
\newpage

\textbf{Simplified-QASM and Conversational Prompt.} 
As observed in the prompt with the QASM example, the token length required for representing quantum circuits can be substantial, especially for systems with a large number of qubits. To address this challenge and facilitate training on resource-constrained hardware such as smaller GPUs, we propose reducing the token length through a process termed \textit{Simplified-QASM}. For example, the sequence "h q[0]; h q[1]; h q[2]; h q[3];" can be compactly represented as "h q[0:4]", merging repetitive instructions into a concise form.

To further enhance the prompt's conversational nature, we append the phrase "The answer is:\textbackslash n" at the end of the Question section. This refinement aligns the prompt with natural language, guiding the model to generate Grover's algorithm probabilities more effectively.
\\
\begin{tcolorbox}[colframe=black!95, colback=white!115, sharp corners]
\textbf{Prompt:}  
\\
Question:  
\\I want you to act as a quantum computer specialized in performing Grover's algorithm.  
I will type a circuit, and you will reply with what a quantum computer should output.  
I want you to only reply with the output in a dictionary that contains the top-30 probabilities and nothing else. The input marked status is: 0000 for a 4-qubit system.  
\\Here is the QASM circuit: \\"h q[0:4]; x q[0:4]; h q[3]; mcx\_0 q[0:4]; h q[3]; x q[0:4]; h q[0:4]; x q[0:4]; h q[3]; mcx\_1 q[0:4]; h q[3]; x q[0:4]; h q[0:4]; x q[0:4]; h q[3]; mcx\_2 q[0:4]; h q[3]; x q[0:4]; h q[0:4]; x q[0:4]; h q[3]; mcx\_3 q[0:4]; h q[3]; x q[0:4]; h q[0:4]; x q[0:4]; h q[3]; mcx\_4 q[0:4]; h q[3]; x q[0:4]; h q[0:4]; x q[0:4]; h q[3]; mcx\_5 q[0:4]; h q[3]; x q[0:4]; h q[0:4];"
\\The answer is:

Answer: \\ 
\{'0000': 0.9596, '0001': 0.0037, '0101': 0.0032, '0100': 0.0032, '0110': 0.0031, '0111': 0.003, '0010': 0.003, '1000': 0.0028, '1100': 0.0028, '1110': 0.0027\}
\end{tcolorbox}

\end{document}